# Interfacial chirality-induced magnetic-field-free switching with high energy efficiency in all-vdW heterostructures

Kai-Xuan Zhang[1,2,3†★$], Suik Cheon[4†], Seungbok Lee[1,2], Joonyoung Choi[5], Jihoon Keum[1,2], Hyuncheol Kim[1,2], Yeochan An[1,2], Woonghee Cho[1,2], Suhan Son[1,2], Jingyuan Cui[1,2§], Pyeongjae Park[1,2¶], Younjung Jo[5], Jun Sung Kim[4], Hyun-Woo Lee[4,6★], and Je-Geun Park[1,2,3★]

[1]Department of Physics and Astronomy, Seoul National University, Seoul 08826, South Korea

[2]Center for Quantum Materials, Department of Physics and Astronomy, Seoul National University, Seoul 08826, South Korea

[3]Institute of Applied Physics, Seoul National University, Seoul 08826, South Korea

[4]Department of Physics, Pohang University of Science and Technology, Pohang 37673, South Korea

[5]Department of Physics, Kyungpook National University, Daegu 41566, South Korea

[6]Asia Pacific Center for Theoretical Physics, Pohang 37673, South Korea

[$]Present address: Department of Physics, Washington University in St. Louis, St. Louis, Missouri 63130, USA

[§]Present address: Department of Physics & Astronomy, University of California, California, 92521, USA

[¶]Present address: Materials Science and Technology Division, Oak Ridge National Laboratory, Oak Ridge, Tennessee 37831, USA

[†] These authors contributed equally to this work.

[★] Corresponding authors: Kai-Xuan Zhang (email: kxzhang.research@gmail.com), Hyun-Woo Lee (email: hwl@postech.ac.kr), and Je-Geun Park (email: jgpark10@snu.ac.kr).

**Abstract:** Chirality, a central concept across many scientific disciplines, continues to inspire the discovery of novel physical phenomena. In condensed matter physics, structural chirality—defined by the absence of mirror plane symmetries—has primarily been explored in *bulk* materials. However, new chiral phenomena can emerge uniquely at the interface, distinct from their bulk counterparts, when a chiral material forms a heterostructure. Here, we demonstrate that all van-der-Waals (vdW) heterostructure composed of the chiral $Co_{1/3}TaS_2$ and the achiral vdW ferromagnet $Fe_3GeTe_2$ exhibits two distinct and unconventional spin-orbit torques originating from the interfacial chirality. These torques enable magnetic-field-free switching of perpendicular magnetization with ultralow current density $\sim 10^6$ A/cm$^2$ and minimal power dissipation $< 10^{15}$ W/m$^3$. Moreover, by replacing $Fe_3GeTe_2$ with a similar vdW ferromagnet, $Fe_3GaTe_2$, but of higher Curie temperature, we achieved the magnetic-field-free switching at room temperature in the $Fe_3GaTe_2/Co_{1/3}TaS_2$ vdW heterostructure. Our findings establish interfacial chirality as a powerful new handle for spintronic control, opening a new pathway to explore chirality-induced phenomena beyond the bulk symmetry constraints – and paving the way toward highly efficient, low-power spintronic devices based on all-vdW heterostructures.

## Introduction

Chiral crystals, defined by the absence of mirror and inversion symmetry, exhibit unique bulk properties that have drawn significant interest in condensed matter physics[1-6]. These bulk chiral effects give rise to exciting new phenomena such as circularly-polarized beam splitting[7], chiral phonons[8], topological transport mediated by chiral Weyl fermions[9], and chirality-induced spin selectivity[10-14]. Moreover, chiral effects in two-dimensional systems have also attracted attention, with studies revealing intriguing effects[15] such as circular dichroism[16], chiral plasmons[17], current-driven collective tuning of helical spin texture[18] and topological spin chirality[19], the nonlinear Hall effect[20-23], and spin transport[24,25]. These studies on chirality are expected to find applications in various fields, including next-generation spintronic, optical, and quantum devices. For instance, chiral materials can exhibit spin selectivity (so-called chirality-induced spin selectivity) without using either magnetism or spin-orbit coupling[26-30]. In spintronics, understanding chiral effects and their utilization is a major challenge. Given this challenge, exploring spin-transport phenomena beyond bulk chiral effects could present an exciting avenue for further research on chirality. In this regard, the interface between heterostructures composed of chiral and achiral materials is an ideal platform for exploring the emergence of novel interfacial phenomena related to chirality.

We investigate the interfacial chiral effect in a bilayer composed of two vdW materials (one of which is a chiral crystal in three dimensions). Two-dimensional magnetic[31-37] and nonmagnetic vdW materials have enormous potential for spintronic device applications[35,38-41] due to their inherent advantages, including high compatibility, tuneability, and flexibility. Notably, efficient stacking with an atomic interface provides an effective approach for spintronic research on interfacial effects in vdW heterostructures[42] and for iterative optimizations. Building on these advantages, vdW heterostructures composed of a chiral crystal and a magnetic achiral vdW material offer a compelling avenue for investigating chiral interface effects and their potential to trigger unconventional spin-orbit torques, ultimately enabling magnetic-field-free magnetization switching.

Magnetic-field-free switching of perpendicular magnetization has been studied in heterostructures where symmetry reduction at the interface plays a crucial role[43,44]. Deterministic switching is demonstrated when the current is applied perpendicular to the vertical mirror symmetry plane, which contains the out-of-plane axis. Still, it vanishes when the current lies within this mirror plane. This typical symmetry configuration corresponds to the $C_{3v}$ point group, where vertical mirror planes restrict the allowed spin-orbit torque components[45,46]. In magnetization switching, chirality corresponds to the complete absence of mirror symmetries, enabling torque components that would otherwise be forbidden. Consequently, in achiral systems with $C_{3v}$ symmetry, magnetic-field-free switching can only occur under limited current directions. By contrast, in a chiral system with $C_3$ symmetry, all mirror symmetries are absent, allowing torque components different from $C_{3v}$ configurations to coexist. The cooperation of these two torque components can enable magnetic-field-free switching beyond the limited current directions allowed in achiral $C_{3v}$ systems. Here, we introduce a magnetic bilayer that contains an exciting chiral van der Waals material[47,48] $Co_{1/3}TaS_2$, which belongs to the space group $P6_322$ and lacks mirror symmetries in all directions[49,50]. In this bilayer, two distinct unconventional spin-orbit torques arise from the interfacial chiral effect (not the bulk chiral effect) and collectively enable magnetisation switching in the absence of a magnetic field. These two torques can lead to a more efficient switching mechanism than previous magnetic-field-free switching experiments, providing a viable pathway for developing high-performance spintronic devices.

In this work, we experimentally demonstrate magnetic-field-free switching driven by chiral all-vdW heterostructures composed of chiral crystal $Co_{1/3}TaS_2$ and a vdW ferromagnet $Fe_3GeTe_2$. Our symmetry analysis reveals two independent unconventional spin-orbit torques arising from interfacial chirality and contributing to magnetic-field-free switching. Furthermore, these torques can cooperate with the proper current direction, leading to low switching current density and minimal switching power dissipation. Additionally, our system exhibits an unconventional current-angle dependence in magnetization switching, distinct from the angular magnetic-field-free switching behavior observed in achiral systems[43,45]. These findings confirm that our magnetic-field-free switching results originate from intrinsic interfacial chirality.

By establishing the feasibility of chiral vdW heterostructures for purely electrical spin-orbit torque switching, our work opens a new avenue for exploring chiral heterostructures for device applications, paving the way for innovative spintronic technologies with enhanced performance and scalability. These results not only expand the material landscape for magnetic-field-free switching but also establish chiral heterostructures as a promising platform for future spintronic devices, offering new opportunities for efficient and high-density memory and logic applications.

## Results

**Magnetic-field-free switching of the $Fe_3GeTe_2/Co_{1/3}TaS_2$ heterostructure.** Figure 1a illustrates the $Fe_3GeTe_2/Co_{1/3}TaS_2$ heterostructure device, a representative case of the ferromagnet/intercalated-transition-metal-dichalcogenide (TMDC) heterostructure. Figure 1b is the typical optical image of the corresponding device and transport measurement schematic. $Fe_3GeTe_2$ is a topological[51] vdW ferromagnet of well-defined Ising-type out-of-plane spin order, displaying a longitudinal resistance ($R_{xx}$) kink near its Curie temperature of ~195 K, accompanied by a resistance upturn at low temperatures (Fig. S1a) and a prominent anomalous Hall effect[52] with transverse resistance ($R_{xy}$) hysteresis loops (Fig. S2a). $Co_{1/3}TaS_2$ is a two-dimensional (2D) antiferromagnet with a space group of $P6_322$, hosting a kink in $R_{xx}$ near its Neel temperature of ~35 K (Fig. S1b) and a normal Hall effect with a linear response of $R_{xy}$ to the external magnetic field (Fig. S2b). Based on our experience and previous reports[39], the nanometer-thin $Fe_3GeTe_2$ nanoflake generally has an $R_{xx}$ resistance of around 100 Ω (Fig. S1a). A $Co_{1/3}TaS_2$ nanoflake with a similar measurement geometry has $R_{xx}$ of around 10 Ω (Fig. S1b). Therefore, most of the current will flow through the $Co_{1/3}TaS_2$ layer, and the electrical signal of the whole heterostructure device primarily originates from this layer. As expected, the $Fe_3GeTe_2/Co_{1/3}TaS_2$ heterostructure also exhibits a kink in $R_{xx}$ at ~26.5K (Fig. 1c) near the Neel temperature of $Co_{1/3}TaS_2$ but far away below the Curie temperature of $Fe_3GeTe_2$. The device also inherits the ferromagnetic anomalous Hall effect, i.e., $R_{xy}$-$H$ hysteresis loops of $Fe_3GeTe_2$ below the Curie temperature (Fig. 1d).

Due to the global inversion symmetry breaking of $Co_{1/3}TaS_2$ from its noncentrosymmetric chiral point group $D_6$, combined with the discrete rotation symmetry breaking by the interface formation, the current may generate a spin polarisation and a resultant spin-orbit torque in the $Fe_3GeTe_2/Co_{1/3}TaS_2$ heterostructure. We observe the magnetic-field-free switching in the $R_{xy}$-Writing current hysteresis loops with zero magnetic field at 75, 90, and 110 K (Fig. 1e). There are several essential details to be noted for the temperature-dependent switching evolution: First, no switching is observed at 50 K within the present current range, probably because the applied current is smaller than the critical switching current. Second, switching emerges at around 10, 9, and 7.5 mA for 75, 90, and 110 K, respectively, with a lower critical switching current at higher temperatures due to reduced magnetic anisotropy. Third, switching vanishes near or above the ferromagnetic Curie temperature as expected. In addition, the magnetic-field-free switching doesn't show up in bare $Fe_3GeTe_2$ nanoflakes (Fig. S3), consistent with the previous investigations on pristine $Fe_3GeTe_2$[39,53,54]. All these experimental features demonstrate that the magnetic-field-free spin-orbit torque switching happens only in the $Fe_3GeTe_2/Co_{1/3}TaS_2$ heterostructure. It also indicates that the magnetic-field-free switching is irrelevant to the antiferromagnetic spin texture of $Co_{1/3}TaS_2$, since the switching appears at a much higher temperature than $Co_{1/3}TaS_2$'s Neel temperature of ~35 K.

We then check the reproducibility of magnetic-field-free switching in another $Fe_3GeTe_2/Co_{1/3}TaS_2$ heterostructure. Figure S4a displays the $R_{xx}$ kink at ~ 26.5 K near the Neel temperature, while Figure S4b shows the ferromagnetic anomalous Hall loops of the entire device. Such a new device also displays magnetic-field-free switching from 75 to 110 K and shares the same switching features as in Fig. 1e. These experimental observations demonstrate the reproducibility of the magnetic-field-free switching in the $Fe_3GeTe_2/Co_{1/3}TaS_2$ heterostructure.

As a passing remark, previous investigations reveal a gigantic intrinsic spin-orbit torque in single $Fe_3GeTe_2$ itself, orders of magnitude larger than that in conventional ferromagnet/heavy-metal composite systems[39]. Consequently, the magnetic memory based on $Fe_3GeTe_2$ is highly energy-efficient, with the switching current density and power dissipation significantly lower than those of the conventional Co/Pt bilayer[53]. However, one significant defect in previous studies is that magnetic-field-free switching cannot be realized merely by current, and a small magnetic field is indispensable for magnetic switching, as shown in our previous works[53,54]. In sharp contrast, our present work has demonstrated a novel, easy way to achieve magnetic-field-free perpendicular switching. More importantly, by replacing $Fe_3GeTe_2$ with a similar vdW ferromagnet $Fe_3GaTe_2$ but of higher Curie temperature, the magnetic-field-free switching is achieved at room temperature in the $Fe_3GaTe_2/Co_{1/3}TaS_2$ vdW heterostructure [Fig. S9].

**vdW chiral interface-induced magnetic-field-free switching.** Next, we perform the symmetry analysis to understand the magnetic-field-free switching in $Fe_3GeTe_2/Co_{1/3}TaS_2$. $Co_{1/3}TaS_2$ belongs to chiral point group $D_6$ (or space group $P6_322$). The point group $D_6$ is characterized by the identity $E$, the six-, three-, and two-fold rotations along the $z$ direction, $C_{6z}$, $C_{3z}$, $C_{2z}$, and twofold rotations along high-symmetry axes (e.g., $x$-axis) $C_{2\parallel}(C_{2x})$ and low-symmetry axes (e.g., $y$-axis) $C_{2\perp}(C_{2y})$ (Fig. 2a). Although the absence of mirror planes in $D_6$ leads to bulk chirality, such bulk chirality alone does not generate unconventional spin-orbit torques required for magnetic-field-free switching[46,55]. To allow the unconventional spin-orbit torque, at least one of the following combinations of symmetries must be broken additionally: (i) $C_{2z}$ and $C_{2\parallel}$ or (ii) $C_{2z}$ and $C_{2\perp}$ (Figs. 2a - 2b). Such symmetry breaking naturally occurs at the interface, particularly in the case of $Fe_3GeTe_2/Co_{1/3}TaS_2$ heterostructure, where $C_{2\parallel}$ and $C_{2\perp}$ symmetries are spontaneously broken at the interface, and the $C_{2z}$ symmetry is additionally broken by the presence of the $Fe_3GeTe_2$ layer, which inherently belongs to the $D_{3h}$ point group without a $C_{2z}$ symmetry. Although the $C_{3z}$ symmetry remains at the interface, it does not hinder the existence of the unconventional spin-orbit torque, but instead gives rise to a three-fold angular dependence of magnetic-field-free switching behaviour (Figs. 2c - 2f). Consequently, at the interface of the $Fe_3GeTe_2/Co_{1/3}TaS_2$ bilayer, symmetry is reduced to the chiral $C_3$ point group by eliminating $C_{2z}$, $C_{2\parallel}$, and $C_{2\perp}$ symmetries, thereby allowing the unconventional spin-orbit torque[46] expressed as follows:

$$T_{3\boldsymbol{m},\perp} = \tau_\perp \boldsymbol{m} \times \left[(m_y E_x + m_x E_y)\hat{x} + (m_x E_x - m_y E_y)\hat{y}\right], (1)$$

$$T_{3\boldsymbol{m},\parallel} = \tau_\parallel \boldsymbol{m} \times \left[(m_x E_x - m_y E_y)\hat{x} + (-m_y E_x - m_x E_y)\hat{y}\right], (2)$$

where $\boldsymbol{m} = (m_x, m_y, m_z)$ is the magnetization, $\boldsymbol{E} = (E_x, E_y, 0)$ is an external electric field, $\hat{x}$ ($\hat{y}$) is the unit direction, and $\tau_{\perp(\parallel)}$ is a torque strength. Here, Eq. (1) represents the unconventional spin-orbit torque that emerges when the simultaneous breaking of (i) occurs, while Eq. (2) corresponds to the torque induced by the breaking of (ii). This simultaneous emergence of two types of unconventional spin-orbit torque is possible due to interfacial chirality. Notably, each torque in Eqs. (1) and (2) share the same mathematical form as the unconventional spin-orbit torque previously reported in achiral $C_{3v}$ systems[45,46]. However, mirror planes are always present in the achiral systems, leading to symmetry-imposed suppression. More precisely, the presence of mirror planes $\sigma_{yz}$ ($\sigma_{zx}$) prohibits Eq. (1) [Eq. (2)], respectively. As a result, only one of the two torques, Eq. (1) or Eq. (2), is permitted in achiral $C_{3v}$ systems. In contrast, in the chiral $C_3$ system ($Fe_3GeTe_2/Co_{1/3}TaS_2$ bilayer), the absence of mirror planes allows both Eq. (1) and Eq. (2) to coexist simultaneously, distinguishing this chiral system from achiral systems[43,45]. Although both can be categorized as low-symmetry systems, the chiral $C_3$ system highlights the unique role of chirality over the achiral $C_{3v}$ system. In the chiral $C_3$ system, the two torques [Eq. (1) and Eq. (2)] can coexist simultaneously, thereby revealing the distinctive contribution of chirality to magnetic-field-free switching. As a result, multiple unconventional spin-orbit torques arising from interfacial chirality can become feasible, providing a significant advantage for realizing field-free switching devices.

Since the two unconventional spin-orbit torques, $T_{3\boldsymbol{m},\perp}$ and $T_{3\boldsymbol{m},\parallel}$ have different angular dependences, they may add up for proper writing current directions to enhance the magnitude of the magnetic-field-free torque $\boldsymbol{T}_{\text{field-free}} = T_{3\boldsymbol{m},\perp} + T_{3\boldsymbol{m},\parallel}$ , leading to low switching current density and low power dissipation. Chirality enables the simultaneous coexistence of two otherwise mutually exclusive torques and their cooperative contribution to magnetic-field-free switching, thereby reducing the critical switching current density. Furthermore, contrary to the $C_{3v}$ system, where magnetic-field-free switching is only permitted on high-symmetry axes and not on low-symmetry axes[45], magnetic-field-free switching in the chiral $C_3$ system ($Fe_3GeTe_2/Co_{1/3}TaS_2$) is achievable on both high- and low-symmetry axes (Figs. 2c - 2f). We want to emphasize that these two results arise from interfacial chirality and are distinct from the previous magnetic-field-free switching in the $C_{3v}$ system[45].

Additionally, we note the change in switching polarity with respect to the direction of $\boldsymbol{E}$. Because there are two kinds of unconventional spin-orbit torque in the $C_3$ system, the angular dependence of magnetic-field-free switching with respect to $\boldsymbol{E} = (cos\ \varphi_{\boldsymbol{E}}, sin\ \varphi_{\boldsymbol{E}}, 0)$ results in a phase shift at $cos\ 3\ \varphi_{\boldsymbol{E}}$. Such a shift is very challenging to measure in an experiment since one cannot know its value precisely. Nevertheless, we can qualitatively explain the change in switching polarity through the symmetry analysis.

Figure 3a illustrates the expected switching from the above symmetry analysis for a general interfacial chiral $C_3$ system. We then check the switching polarity change by experiment. Figure 3b shows the optical image of an $Fe_3GeTe_2/Co_{1/3}TaS_2$ device with twelve electrodes at every 30° rotated from each other. As expected, we indeed observe the three-fold character of the magnetic-field-free switching with switching polarity changing every 60° (Fig. 3c). Magnetic-field-free switching also exists at 90° rather than vanishing, indicating that the interfacial three-fold rotation symmetry along the $z$ direction is perturbed by complex interactions in real devices: the two unconventional spin-orbit torques combine to push away the no-switching angle to a direction different from all high-symmetry directions. A similar result is obtained in another device in Fig. S5. In addition, we used focused ion beam (FIB) etching to fabricate two devices with sun-shaped geometry, where each electrode was separated by a FIB-etched region, so that the current would flow along a specifically designed direction without mutual interference[56,57]. The switching polarity results for sun-shaped devices resemble those for pristine heterostructure devices [Fig. S11 and Fig. S12]. This observation indicates that the electrical field and the consequent current flow are mainly determined by the designed electrodes, even in pristine samples without etching for our systems, and that this switching polarity behaviour also remains consistent with the symmetry-based analysis. All these experimental observations are consistent with our symmetry analyses. In short, both theoretically and experimentally, we demonstrate that the chiral vdW interface can be important in activating bulk-symmetry-forbidden magnetic-field-free switching by current-driven spin-orbit torque.

**Magnetic-field-free switching by current with high efficiency.** Figure 4 shows the final $Fe_3GeTe_2/Co_{1/3}TaS_2$ device, in which we performed various switching measurements with well-behaved performance. Figure 4a represents the time sequence of applied writing current and consequent Hall response, where the Hall resistance $R_{xy}$'s switching is readily controlled by the writing currents. Figure 4b shows more prominent magnetic-field-free switching over a broader temperature range with a lower critical switching current than in the abovementioned devices. In Fig. 4c, magnetic-field-free switching was measured repeatedly with four different cases: Case 1 ($H_z$: 0.4 T initialization → 0 T; $I$: 10 → −10 → 10 mA), Case 2 ($H_z$: 0 T no initialization; $I$: 10 → −10 → 10 mA), Case 3 ($H_z$: −0.4 T initialization → 0 T; $I$: 10 → −10 → 10 mA), Case 4 ($H_z$: −0.4 T initialization → 0 T; $I$: −10 → 10 → −10 mA). These switching loops in four different cases reasonably overlap with one another, demonstrating reliable switching performance. Figure 4d summarizes the switching current density and power dissipation in our $Fe_3GeTe_2/Co_{1/3}TaS_2$ system and other recent magnetic-field-free switching systems[43,45,58-60]. Details of how to estimate the switching current density are provided in the Supporting Note 3. Our work demonstrates comparable or lower switching current density and power dissipation compared to those previously reported novel systems[43,45,58-60].

Multiple unconventional spin-orbit torques arising from the interfacial chirality can become feasible and provide a significant advantage in realizing field-free switching devices. In our system, the two unconventional spin-orbit torques, $T_{3\boldsymbol{m},\perp}$ and $T_{3\boldsymbol{m},\parallel}$ can coexist and jointly contribute to the magnetic-field-free torque $\boldsymbol{T}_{\text{field-free}}=T_{3\boldsymbol{m},\perp}+T_{3\boldsymbol{m},\parallel}$, providing a possible explanation for the low switching current density. In addition, we estimated $|\Delta H_{3\text{m},\perp}|/J \sim 8.7\times10^{-12}\ \text{TA}^{-1}\text{m}^2$ and $|\Delta H_{3\text{m},\parallel}|/J \sim 7.3\times10^{-12}\ \text{TA}^{-1}\text{m}^2$ from a second-harmonic measurement experiment (Supporting Note 2; Fig. S10), which is on the same order as the record-high unconventional torque efficiency of $TaIrTe_4/Fe_3GaTe_2$ system (e.g., $|H_{z,\text{DL}}|/J \sim 6.78\times10^{-12}\ \text{TA}^{-1}\text{m}^2$ in Ref.[61]). Such high torque efficiencies are more important factors for causing low switching current density, which may be facilitated by possible topological bands and Berry curvature hotspots.

## Discussion

Magnetic-field-free switching of perpendicular magnetization is highly desired for integrating spin-orbit torque devices. However, previous demonstrations of field-free switching have suffered from the presence of mirror-symmetry planes. A classic example is the achiral $L1_1$-ordered CuPt/CoPt bilayer[45], where three vertical mirror planes are mutually related to each other through the three-fold rotation symmetry. No deterministic switching is allowed for that system when the current is aligned parallel to one of these mirror planes. In the achiral $WTe_2/Fe_3GeTe_2$ bilayer[43], where there is one vertical mirror plane (e.g., the $zx$ plane or $yz$ plane in Fig. 2a) with two-fold rotation symmetry, deterministic switching is similarly forbidden when the current is aligned parallel to the mirror plane. In this respect, our chiral bilayer $Co_{1/3}TaS_2/Fe_3GeTe_2$ is special since no mirror symmetry plane prevents field-free switching. Notably, most bulk chiral materials do not play a significant role in generating unconventional spin-orbit torques responsible for field-free switching[46,55], as these effects primarily originate from the chiral interface. However, the absence of the mirror planes cannot remove the special current direction along which the field-free switching does not occur, since the existence of such special directions is guaranteed by the linearity of the spin-orbit torque itself (Fig. S6). Instead, the advantage of the chiral bilayer is the simultaneous existence of multiple unconventional spin-orbit torques [Eqs. (1) and (2)], whose cooperation may enhance the efficiency of the field-free switching and shift the zero-switching axis away from principal crystalline directions. We remark that our efficiency is also higher than that of the reported achiral bilayer system $TaIrTe_4/Fe_3GaTe_2$[59] and achiral trilayer system $PtTe_2/WTe_2/CoFeB$[60]. In addition, we have summarized the significance and novelty of this work in the Supporting Note 1, thereby better conveying the uniqueness and distinctions of the present work.

In summary, we demonstrate magnetic-field-free magnetization switching enabled by interfacial chirality in all vdW $Fe_3GeTe_2/Co_{1/3}TaS_2$ heterostructures. Our results establish interfacial chirality as a promising route to control magnetization in 2D vdW systems. By uncovering chirality-driven interfacial phenomena beyond the constraints of bulk symmetry, we reveal a new class of unconventional spin-orbit torque driven solely by electrical current, with exceptional energy efficiency. Leveraging the vast family of intercalated chiral TMDCs, this approach opens a broad design space for engineering chiral interfaces in vdW heterostructures. Our work presents a powerful materials strategy for realizing magnetic-field-free switching platforms, offering a scalable, energy-efficient path towards next-generation spin-orbit torque memory technologies and their industrial deployment.

## Methods

**Single-crystal growth and device fabrication.** $Fe_3GeTe_2$ single crystals were grown using the chemical vapour transport method[62,63]. $Co_{1/3}TaS_2$ single crystals were grown via a combined solid-state reaction and chemical vapour transport method[49,50]. First, polycrystalline $Co_{1/3}TaS_2$ was made by the solid-state reaction to make the sample chemically homogeneous: High-purity Co, Ta, and S powders were sealed in a vacuum quartz tube with a ratio of 1.18:3:6, and then the quartz tube was placed in a furnace at 900 ˚C for sintering for about one week. Second, the obtained polycrystals were ground to fine powders and sealed in a new vacuum quartz tube with iodine as the transport agent. This new tube was placed in a two-zone furnace with a temperature gradient of 940 ˚C (source) to 860 ˚C (sink) for ten days to eventually grow the single crystals through the chemical vapour transport process.

$Co_{1/3}TaS_2$ nanoflakes were exfoliated onto the $SiO_2$/Si substrate by a conventional mechanical exfoliation method with Scotch tape from the above as-grown single crystals. $Fe_3GeTe_2$ nanoflakes were mechanically exfoliated onto the $SiO_2$/Si substrate and then dry-transferred onto the pre-prepared $Co_{1/3}TaS_2$ nanoflakes by the polycaprolactone (PCL) method[64,65]. The entire process was carried out in a glove box filled with Argon gas. The samples were placed in a load-lock box within the glove box, then removed. Immediately, PMMA polymer was spin-coated onto the samples and then baked at 130 ˚C for 1.5 min, followed by a standard electron beam lithography (EBL). Finally, 80/10 nm Au/Ti electrodes were deposited onto the device by electron-beam evaporation under high vacuum.

**Electrical transport measurements.** We performed transport measurements using a resistivity probe operated within a cryostat. The writing current was applied by a Keithley 6220 or 6221, and the Hall resistance was measured using a standard lock-in technique or a Keithley 2182A. A gold wire connected the electronic chip to the sample's electrodes via wire bonding. An antistatic wrist strap was used whenever one touched the electrical measurement system to prevent possible damage from electrostatic discharges or shocks to the sample during operation. For the $R_{xy}$-$H$ curves, we applied an out-of-plane magnetic field perpendicular to the sample surface unless otherwise stated.

**Data availability.** Relevant data supporting the key findings of this study are available within the Article and the Supplementary Information. The raw data generated during the current study are available from the corresponding authors upon request.

**Acknowledgements**

One of the authors (J.G.P.) acknowledges the hospitality of the Indian Institute of Science, where parts of the manuscript were prepared, and the generous financial support of the Infosys Foundation. The Samsung Advanced Institute of Technology also supported this work at both SNU and POSTECH.

**Funding**

The work at CQM and SNU was supported by the Samsung Science & Technology Foundation (Grant No. SSTF-BA2101-05), the Leading Researcher Program of the National Research Foundation of Korea (Grant Nos. 2020R1A3B2079375 and RS-2020-NR049405), and the National Research Foundation (NRF) of Korea (Grant No. RS-2026-25484445), funded by the Ministry of Science and ICT. The theoretical works at POSTECH were funded by the National Research Foundation (NRF) of Korea (Grant Nos. 2020R1A2C2013484 and RS-2024-00410027).

**Author contributions**

K.X.Z., J.G.P., and H.W.L. conceived the project. K.X.Z., Y.A., W.C., P.P., and J.S.K. grew the $Fe_3GeTe_2$, $Co_{1/3}TaS_2$, and $Fe_3GaTe_2$ single crystals. K.X.Z. fabricated the nanodevices with help from S.L.. K.X.Z. performed the transport experiments assisted by J.K., H.K., J.C. and S.S., J.C., and Y.J.. S.C. and H.W.L. performed the theoretical calculations and analyses. K.X.Z., S.C., H.W.L., and J.G.P. analyzed the data and wrote the manuscript with contributions from others.

## Competing interests

The authors declare no competing interests.

## Figures

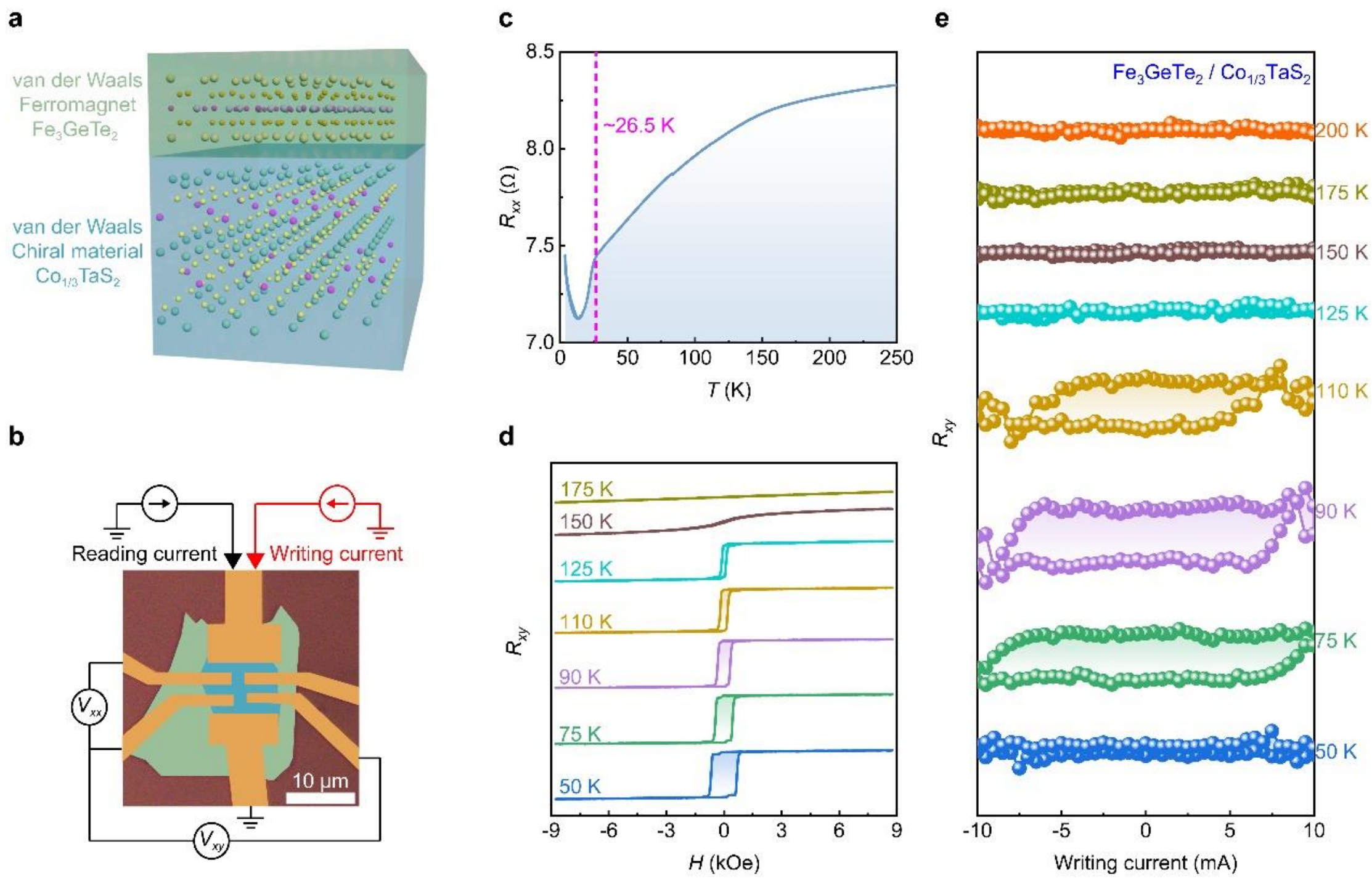

**Fig. 1 | Magnetic-field-free switching demonstration of $Fe_3GeTe_2/Co_{1/3}TaS_2$ heterostructure.** **a**, Illustration of $Fe_3GeTe_2/Co_{1/3}TaS_2$ heterostructure device. **b**, Typical optical image of the $Fe_3GeTe_2/Co_{1/3}TaS_2$ device and measurement schematic. The white scale bar is 10 µm. **c**, Temperature-dependent longitudinal resistance $R_{xx}$ of the $Fe_3GeTe_2/Co_{1/3}TaS_2$ device. The pink dashed line indicates the resistance kink at 26.5 K near the Neel temperature $T_{\mathrm{Neel}}$ of $Co_{1/3}TaS_2$. **d**, Transverse Hall resistance $R_{xy}$ as a function of magnetic field for the $Fe_3GeTe_2/Co_{1/3}TaS_2$ device at various temperatures. The entire device hosts well-behaved ferromagnetic behaviour with a clear anomalous Hall effect. **e**, $R_{xy}$ as a function of writing current with zero magnetic field at various temperatures for the $Fe_3GeTe_2/Co_{1/3}TaS_2$ heterostructure. Magnetic-field-free switching occurs at 75, 90, and 110 K, with a hysteresis loop and a gradually decreased critical switching current.

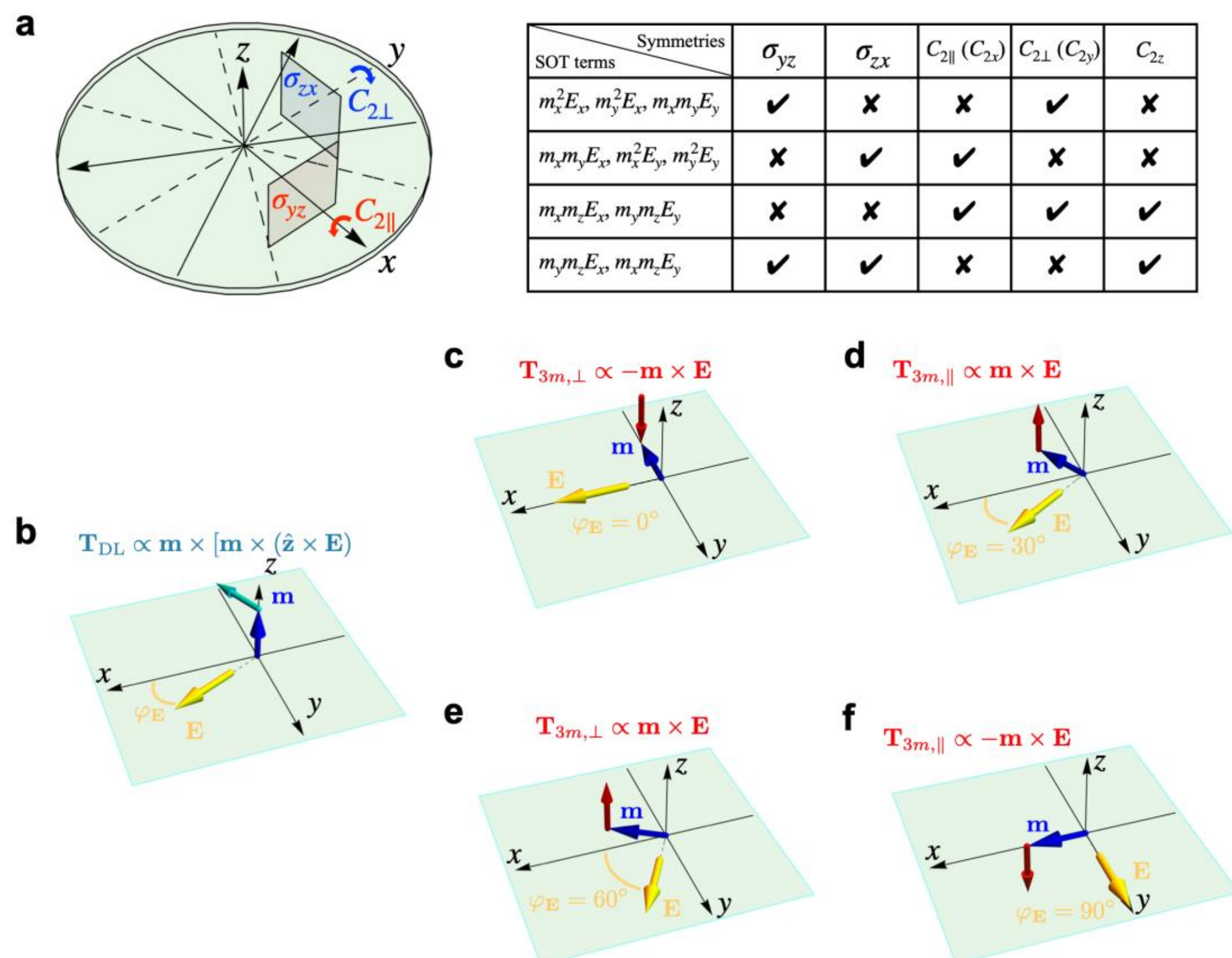

| SOT terms \ Symmetries | $\sigma_{yz}$ | $\sigma_{zx}$ | $C_{2\parallel}$ ($C_{2x}$) | $C_{2\perp}$ ($C_{2y}$) | $C_{2z}$ |
|---|---|---|---|---|---|
| $m_x^2E_x$, $m_y^2E_x$, $m_xm_yE_y$ | ✔ | ✘ | ✘ | ✔ | ✘ |
| $m_xm_yE_x$, $m_x^2E_y$, $m_y^2E_y$ | ✘ | ✔ | ✔ | ✘ | ✘ |
| $m_xm_zE_x$, $m_ym_zE_y$ | ✘ | ✘ | ✔ | ✔ | ✔ |
| $m_ym_zE_x$, $m_xm_zE_y$ | ✔ | ✔ | ✘ | ✘ | ✔ |

**Fig. 2 | Interface symmetry analysis and schematic illustration of unconventional spin-orbit torque for magnetic-field-free switching. a**, Symmetry analysis for the out-of-plane spin torque (z component of the spin torque). "✔" ("✘") represents that the torque term is symmetry allowed (forbidden). In the table, the first two lines correspond to magnetic-field-free switching related unconventional damping-like spin-orbit torque terms, where the first and second lines represent the components of $\boldsymbol{T}_{3m,\perp}$ [Eq. (1)] and $\boldsymbol{T}_{3m,\parallel}$ [Eq. (2)], respectively. The last two lines correspond to conventional damping-like spin-orbit torque terms. Here, $\boldsymbol{m} = (m_x, m_y, m_z)$ is the magnetization, $\boldsymbol{E} = (E_x, E_y, 0)$ is an external electric field, $\sigma_{yz}$ ($\sigma_{zx}$) denotes the mirror symmetry with respect to $yz$ ($zx$) plane, and $C_{2i}$ is the two-fold rotation along the $i$-direction. **b,** At the initial state ($\boldsymbol{m} \parallel \hat{\boldsymbol{z}}$), the conventional damping-like

spin-orbit torque [$\boldsymbol{T}_{\mathrm{DL}} \parallel (\sin\varphi_{\boldsymbol{E}}, -\cos\varphi_{\boldsymbol{E}}, 0)$] exists alone. **c-f**, When $\boldsymbol{m}$ is placed on the in-plane due to the conventional damping-like spin-orbit torque, the unconventional spin-orbit torques act as the out-of-plane torque, allowing deterministic switching. **c** (**e**), $\boldsymbol{T}_{3m,\perp}$ torque direction is $-\hat{z}$ $(+\hat{z})$ for $\varphi_{\boldsymbol{E}} = 0°$ $(\varphi_{\boldsymbol{E}} = 60°)$. And $\boldsymbol{T}_{3m,\perp}$ vanishes for $\varphi_{\boldsymbol{E}} = 30°$ and $\varphi_{\boldsymbol{E}} = 90°$. **d** (**f**), $\boldsymbol{T}_{3m,\parallel}$ torque direction is $+\hat{z}$ $(-\hat{z})$ for $\varphi_{\boldsymbol{E}} = 30°$ $(\varphi_{\boldsymbol{E}} = 90°)$. Also, $\boldsymbol{T}_{3m,\parallel}$ vanishes for $\varphi_{\boldsymbol{E}} = 0°$ and $\varphi_{\boldsymbol{E}} = 60°$. Here, the blue arrow denotes magnetization, the yellow arrow denotes the electric field direction, and the red arrow denotes out-of-plane unconventional spin-orbit torque.

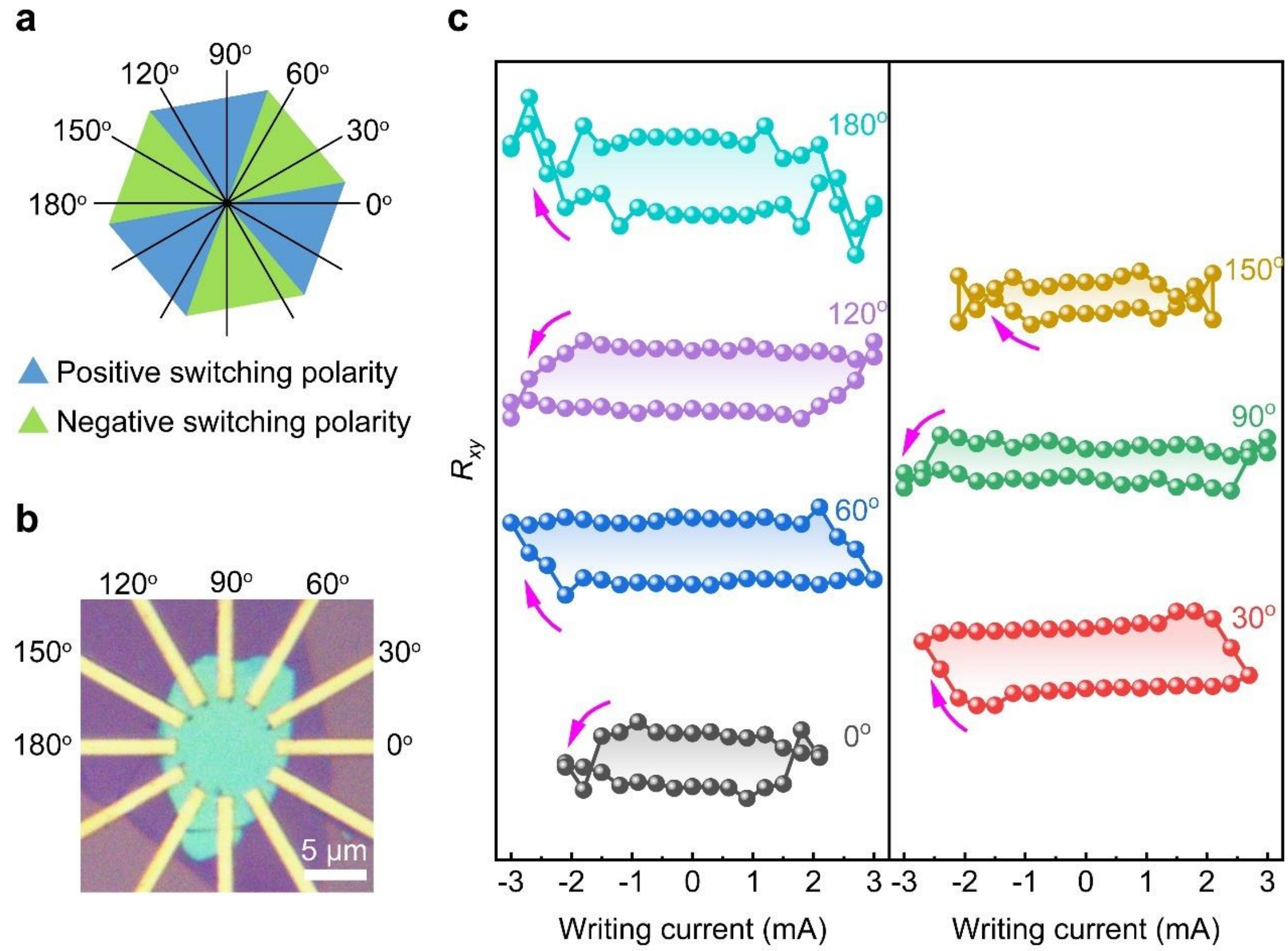

**Fig. 3 | Current direction dependence of the magnetic-field-free switching. a**, Expected three-fold switching polarity by the chiral interface $C_3$ symmetry. **b**, Optical image of the $Fe_3GeTe_2/Co_{1/3}TaS_2$ heterostructure, where the writing current direction is rotated every 30° as illustrated. **c**, Magnetic-field-free switching at each angle. The switching polarity changes every 60° as indicated by the red curved arrows, exhibiting a three-fold dependence.

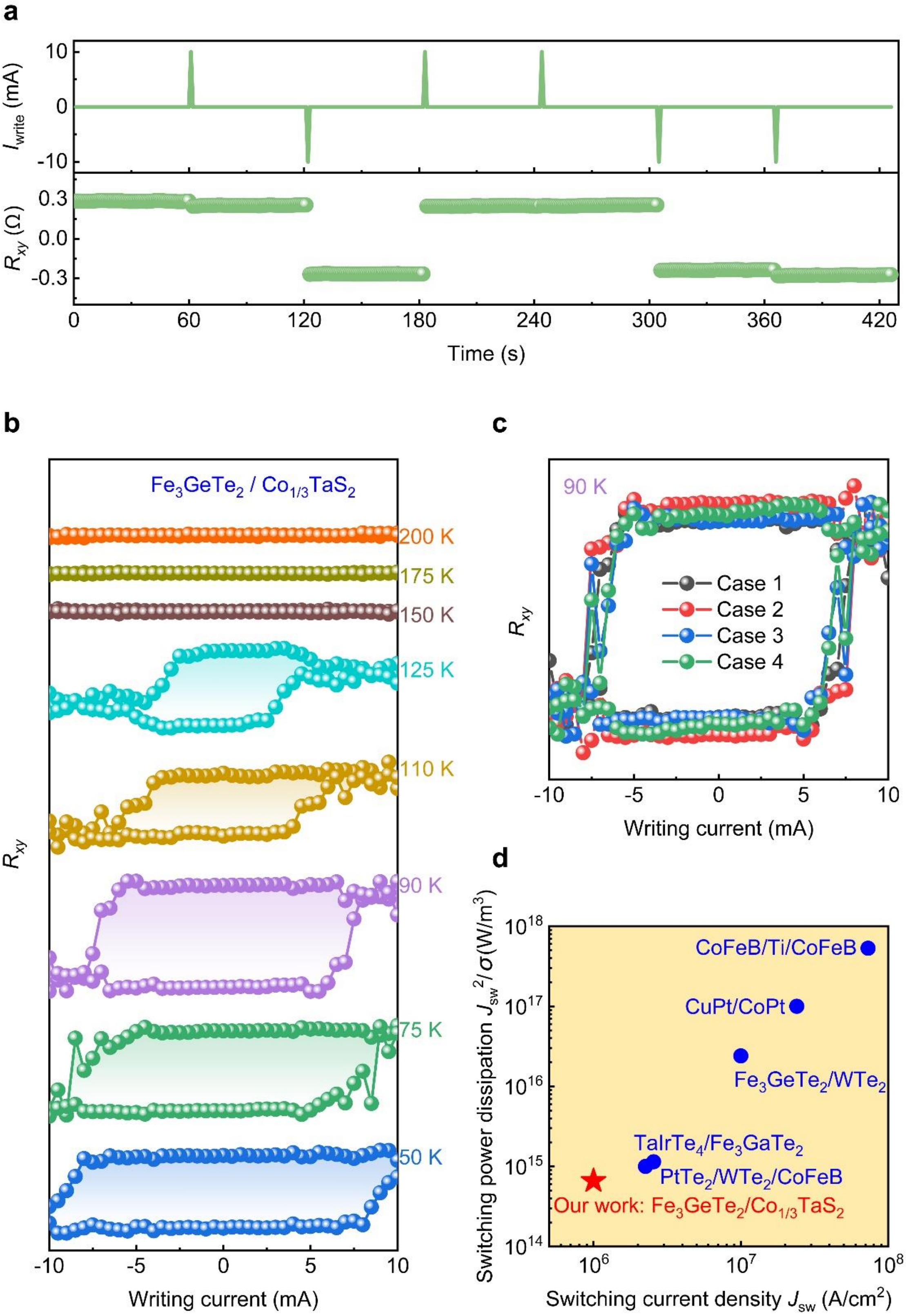

**Fig. 4 | Well-behaved switching performance readily controlled by writing current with high efficiency. a**, Time sequences of applied writing current and responsive $R_{xy}$ at 75 K of a $Fe_3GeTe_2$/$Co_{1/3}TaS_2$ heterostructure. The Hall resistance $R_{xy}$ is well

controlled by writing current directions. **b**, $R_{xy}$-Writing current curves with no magnetic field at various temperatures. Magnetic-field-free switching in this device occurs in a broader temperature range of 50-125 K, much more significantly. **c**, Magnetic-field-free switching loops at 90 K for four different cases: Case 1 ($H_z$: 0.4 T initialization → 0 T; $I$: 10 → −10 → 10 mA), Case 2 ($H_z$: 0 T no initialization; $I$: 10 → −10 → 10 mA), Case 3 ($H_z$: −0.4 T initialization → 0 T; $I$: 10 → −10 → 10 mA), Case 4 ($H_z$: −0.4 T initialization → 0 T; $I$: −10 → 10 → −10 mA). These switching loops are nearly overlapped, demonstrating reliable magnetic-field-free switching performance. **d**, Summary of the switching current density $J_{sw}$ and power dissipation $J_{sw}^2/\sigma$ ($\sigma$ is the conductivity). Details of how to estimate the switching current density are provided in the Supporting Note 3. Our work demonstrates comparable or lower switching current density and power dissipation to those of recent novel magnetic-field-free switching systems[43,45,58-60].